\documentclass[review,final]{elsarticle}

\journal{Journal of Sound and Vibration}

\usepackage[mathscr]{eucal}
\usepackage{amsmath}
\usepackage{amssymb}
\usepackage{mathtools}
\usepackage{amsthm}
\usepackage{times}
\usepackage[utf8]{inputenc}
\usepackage[english,russian]{babel}
\usepackage[T2A]{fontenc}
\usepackage{setspace}
\usepackage{url}
\usepackage[colorlinks,allcolors=blue,unicode,
 pdftitle={The anti-localization of non-stationary linear waves}]{hyperref}
\usepackage{afterpage}
\usepackage[svgnames]{xcolor}

\def\d{\mathrm d}

\def\prpr#1{#1^{\prime\prime}}

\def\sign{\operatorname {sign}}

\def\sign{\operatorname{sign}}

\def\Re{\operatorname{Re}}

\def\I{\mathrm i}

\def\cc{\mathrm{c.c.}}
\renewcommand{\=}{\stackrel{\mbox{\scriptsize def}}{=}}
\def\EXP#1{\mathrm e^{#1}}
\def\Res{\operatorname{Res}}

\theoremstyle{remark}
\newtheorem{remark}{Remark}
\theoremstyle{theorem}
\newtheorem{theorem}{Theorem}

\begin{document}
\selectlanguage{english}
\begin{frontmatter}
\title{The anti-localization of non-stationary
linear waves and its relation to the localization. The simplest
illustrative problem}

\author[ipme]{Ekaterina V.~Shishkina}
\ead{shishkina\_k@mail.ru}
\author[ipme]{Serge N.~Gavrilov\corref{mycorrespondingauthor}} 
\ead{serge@pdmi.ras.ru}
\cortext[mycorrespondingauthor]{Corresponding author}
\author[ipme]{Yulia A.~Mochalova}
\ead{yumochalova@yandex.ru}
\address[ipme]{Institute for Problems in Mechanical Engineering RAS, V.O., Bolshoy
pr. 61, St.~Petersburg, 199178, Russia}
\begin{abstract}	
{We introduce a new wave phenomenon, which can be observed in continuum and
discrete systems, where a
trapped mode exists under certain conditions, namely, 
the anti-localization of non-stationary linear waves.} 
This is zeroing of the non-localized propagating component of the wave-field
in a neighbourhood of an inclusion. 
In other words, it is a tendency for
non-stationary waves to propagate avoiding a neighbourhood of an inclusion.
The anti-localization is caused by a
destructive interference of the 
harmonics
involved into the representation of the solution in the form of a Fourier
integral.
The anti-localization is associated 
with the waves from the pass-band, whereas the localization related with a
trapped mode
is due to poles inside the stop-band.
In 
the framework of a simple illustrative problem considered in the paper, we
have
demonstrated that the anti-localization exists for all cases excepting the
boundary of the domain in the parameter space where the wave
localization occurs.
Thus, the anti-localization can be
observed in the absence of the localization as well as together with the
localization.
We also investigate the influence of the anti-localization on
the wave-field in whole.
\end{abstract}
\begin{keyword}	
anti-localization
\sep
localization
\sep
trapped mode
\sep
non-stationary waves
\sep
linear waves
\sep
vibration
\sep
inclusion
\sep
defect
\end{keyword}

\end{frontmatter}

\section{Introduction}

It is well known that in an infinite linear (continuum or discrete) almost
homogeneous system involving  
a finite number 
of inclusions or defects, {provided that
there is a stop-band in the dispersion characteristics for the corresponding pure
homogeneous system},
one can observe the linear wave localization ({see, e.g., studies
\cite{Kuznetsov2002,Ind-book-R2E,Andrianov2012,porter2007trapped,Mishuris2020} and references there}).
This type of the wave localization is related to the formation of a discrete
part of spectrum of natural frequencies inside a stop-band under certain conditions, which are
fulfilled in some domain of the parameter space (we call this {\it the localization
domain}). 
In continuum mechanics the corresponding localized modes are known 
as the trapped (or trapping) modes, which have been first time discovered 
by Ursell \cite{ursell1951trapping} in the theory of
surface water waves. 

In discrete
mechanical systems the analogous phenomenon, to the best of our knowledge, was
first time
described by Montroll and Potts \cite{Montroll1955}, {though 
it was previously known in physics for non-mechanical systems \cite{Conwell1950,Koster1954a,Koster1954}.}
According to Luongo \cite{luongo2001mode,Luongo1992}, in physics this type of
localization is known as the strong localization. The presence of frequencies
inside the stop-band leads to the possibility to localize non-stationary
waves, i.e., to trap some portions of the wave energy forever near
inhomogeneities (in the absence of dissipation). 
One can observe undamped localized vibration of an infinite system subjected
to an impulse
loading.
For a discrete mechanical system this was shown first time by
Teramoto \& Tokeno \cite{Teramoto1960}. 
For a continuum system Ursell declared in 1987 \cite{Ursell1987}
that this fact had not been demonstrated, though, in
reality, Kaplunov showed 
\cite{kaplunov1986torsional}
it in 1986 not knowing that the considered system  
possesses a trapped mode. The latter fact was discovered later \cite{Abramyan1994,kaplunov1995simple}.
Nowadays, the localization of 
non-stationary waves in continuum systems is described in many studies 
\cite{Ind-book-R2E,Mishuris2020,mciver2003excitation,Indeitsev2006,indeitsev2012motion,gavrilov2002etm,Gavrilov2019nody,Gavrilov2022jsv,indeitsev2016evolution,Shishkina2019jsv}.

Trapped modes characterized by natural frequencies inside a stop-band should
be distinguished from so-called embedded trapped modes, see, e.g., review 
\cite{Linton2007}. The latter ones are
characterized by the discrete spectrum of natural frequencies embedded into
the continuous spectrum (i.e., into the pass-band). 
The localization associated
with the embedded trapped modes is not considered in this paper.

The present paper demonstrates that in the discussed above class of systems
{(continuum or discrete)}, 
where we can 
expect the wave localization, 
a new wave phenomenon can be generally observed. We suggest calling
it {\it the anti-localization of non-stationary linear waves}. 
This is zeroing of the non-localized propagating component of
the wave-field
in a neighbourhood 
of an inclusion.
In other words, it is a tendency for non-stationary waves to propagate avoiding 
a neighbourhood of an inclusion. The anti-localization is caused by a destructive interference of the propagating harmonics
involved into the representation of the solution in the form of a Fourier
integral.
{The anti-localization is associated 
with the waves from the pass-band (including the cut-off frequency
separating 
the pass-band and the stop-band), whereas the corresponding localization 
is due to poles inside the stop-band.}
In Sect.~\ref{sec-problem} we introduce the simplest illustrative problem 
to demonstrate what
the anti-localization of non-stationary waves is, the influence of the
anti-localization to the wave-field in whole, 
and the influence of the wave localization to the anti-localization.
We show that for the problem under consideration 
the anti-localization of non-stationary waves exists 
in all cases excepting the
boundary of the localization domain. Thus, the anti-localization can be
observed in the absence of the localization as well as together with the
localization.

{The mechanical system we deal with {(an 
infinite taut string on the Winkler foundation equipped with a discrete
mass-spring oscillator)} can be considered as an extension of
the system studied in \cite{kaplunov1986torsional}.
Note that the identical mechanical system was previously considered in studies 
\cite{Glushkov2011a,Gavrilov2019asm}, where the anti-localization
was not discovered.}
Our results are in agreement with
observations in studies
\cite{,kaplunov1986torsional,hemmer1959dynamic,Mueller1962,Kashiwamura1962,Rubin_1963,Mueller2012},
which deal with non-stationary oscillation caused by an impulse
source at an inclusion in (discrete or continuum) 
systems where the localization of non-stationary
waves is possible\footnote{In the discrete case it is more correct to
speak about quasi-waves, since the perturbations propagate at an infinite
speed}; see more details in Sect.~\ref{sec-discussion} (Discussion).


The term ``anti-localization'' {in the sense we use it in the paper} was introduced by
Shishkina \& Gavrilov in recent study~\cite{arxiv2206.08079}, though the
term ``weak anti-localization'' is commonly known in modern {quantum
physics}. According to \cite{Janssen2001}, the weak
anti-localization is a phenomenon observable in disordered systems, 
which has been predicted
in \cite{Hikami1980}. The term ``weak anti-localization'', as far as we know, 
has been suggested in \cite{Bergmann1982} as a phenomenon opposite to the weak
localization predicted in 
\cite{Abrahams1979,Gorkov1979} (see also \cite{Gorkov_1996}). The latter one
is a
{spatially localized} amplification
of a stationary wave-field
composed
of the propagating waves from the pass-band
\cite{luongo2001mode,Luongo1992,Pierre1990}.
The weak
localization is observable {\it only} in disordered systems 
\cite{luongo2001mode,Janssen2001}. 
It emerges 
due to a constructive wave 
interference at some inhomogeneities, whereas
the weak anti-localization is caused by a destructive interference.
In optics \cite{Akkermans1986} and acoustics \cite{Bayer1993,Larose_2004,Tourin1997} the weak localization is known as the coherent back-scattering.
At the same time, we have not found
any study on the weak anti-localization in acoustics or wave mechanics,
although there are may be some.
Note that the strong localization, which is related with waves from the stop-band,
can also be observed in disordered systems, see the Anderson localization 
\cite{Anderson1958,Hodges1982,Kissel1988,Pierre1990}. 

Thus, the anti-localization of non-stationary waves discussed
in this paper
and the weak anti-localization, which is known
in quantum physics, have different nature.
Indeed, the latter one is observable only in disordered systems,
whereas we consider ordered deterministic systems. 
At
the same time, the meaning of the term ``anti-localization'' is the same as in
quantum physics. In both cases one can see zeroing (or, strictly speaking,
asymptotic weakening) of the wave-field near 
some inhomogeneities. 

One more physical phenomenon, which should be distinguished from
the anti-localization of non-stationary waves and, therefore, should be
referenced here, is the
blocking of running waves
\cite{Babeshko1992,Alves1995,Glushkov2006,Glushkov2006a,Glushkov2015,Glushkov2015a,Glushkov2018}. The blocking is observable in diffraction problems at
resonant (or ``almost''  resonant) frequencies 
in systems where an embedded trapped mode can exist
under certain conditions. Thus, the blocking is a stationary phenomenon, which
is beyond the scope of our paper.

{Note that recent paper \cite{Ying2021} introduces the term
``anti-localization'' while considering an ordered finite non-linear mechanical system.
The term again means zeroing of the wave-field
near an inhomogeneity. Since an ordered
mechanical system is under consideration, the phenomenon is not a weak
anti-localization. Also, it is not an anti-localization of non-stationary waves,
since the stationary deterministic vibration is under consideration.}

\section{The illustrative problem and its solution}
\def\xi{x}
\def\tau{t}
\label{sec-problem}
We consider transverse oscillation of an infinite taut string on the Winkler
elastic foundation. The string is equipped with a discrete mass-spring
oscillator, which is subjected to an impulse loading.
The governing equation
in the dimensionless form is
\begin{gather}
\prpr u-\ddot u
-u=\big(M\ddot u+Ku-\delta(\tau)\big)\,\delta(\xi).
\label{OSC-maineq-SPRING}
\end{gather}
Here and in what follows, we denote by prime the derivative
with respect to spatial coordinate $\xi$ and
by overdot the derivative with respect to
time $\tau$, non-negative constants $M$ and $K$ (the problem parameters) are
the dimensionless mass and the stiffness characterizing the oscillator.  Zero
initial conditions are assumed. 

The solution can be represented in the form of the following Fourier integral:
\begin{multline}
u(\xi,\tau)=
\frac{1}{2\pi}\int_{-\infty}^{+\infty}  
\mathscr G(x,\Omega)
\exp(-\I \Omega \tau)\, \d\Omega
=
\frac{1}{2\pi}\int_{0}^{\Omega_\ast}  
\mathscr G^{\mathrm{stop}}(x,\Omega)
\exp(-\I \Omega \tau)\, \d\Omega
\\+
\frac{1}{2\pi}\int_{\Omega_\ast}^\infty  
\mathscr G^{\mathrm{pass}}(x,\Omega)
\exp(-\I \Omega \tau)\, \d\Omega
+\cc=
I^{\mathrm{stop}}+I^{\mathrm{pass}}+\cc
\label{u-exact}
\end{multline}
%
Here $\mathscr G(x,\Omega)$ is the corresponding Green function in the
frequency domain \cite{Gavrilov2022jsv}:
\begin{align}
&\mathscr G(\xi, \Omega)=\mathscr G^{\mathrm{stop}}(\xi, \Omega)\=\frac{\exp(-\sqrt{1-\Omega^2}|\xi|)}
{2\sqrt{1-\Omega^2}+K-M\Omega^2},
\quad
\Omega\in\mathbb S
;
\\
&\mathscr G(\xi,\Omega)=\mathscr G^{\mathrm{pass}}(\xi, \Omega)\=-\frac{\exp(\I\sqrt{\Omega^2-1}|\xi|\sign
(\Omega))}{2\I\sign(\Omega) \sqrt{\Omega^2-1}-K+M\Omega^2},
\quad
\Omega\in\mathbb P
.
\label{Green-function-K=0}
\end{align}
Here 
$\mathbb S=(-\Omega_\ast,\Omega_\ast)$ and $\mathbb
P=(-\infty,-\Omega_\ast)\cup(\Omega_\ast,\infty)$ are the stop-band and the
pass-band, respectively; $\Omega_\ast=1$ is the cut-off frequency, which
separates the bands. 

The integral $I^{\mathrm{stop}}$ describes, in particular, a localized non-vanishing
oscillation that one can observe in the system under certain conditions.
Namely, in the case $K>M$ a rough estimate $I^{\mathrm{stop}}=O(t^{-1})$
$(t\to\infty)$ 
can be obtained 
(the details of the mathematical
technique can be
found in \ref{App-B}).
In the case 
\begin{equation}
K<M
\label{loc-domain}
\end{equation}
a localized (trapped) mode exists in the system
\cite{Ind-book-R2E,Gavrilov2022jsv,Gavrilov2019asm}.
Inequality 
\eqref{loc-domain}
defines the localization domain in the 2D parameter
space.
If this inequality is true, in the interval $(0,\Omega_\ast)$ there exists a
{simple root~$\Omega_0$} of the denominator for $\mathscr G^{\mathrm{stop}}$ such that
\begin{equation}
\Omega_0^2{}=\frac 2{M^2}\left(
\sqrt{M^2-MK+1}+\frac{MK}2-1
\right),
\label{OSC-fr-root}
\end{equation}
 and,
therefore,
integral $I^{\mathrm{stop}}$ does not exist in the classical
sense.
In the
latter case,
integral $I^{\mathrm{stop}}$ 
should be considered as the Fourier transform for
a generalized function. To regularize this {we can apply} 
\cite{Gavrilov1999jsv,arxiv2206.08079}
the limit absorption
principle. {Finally, for $t\to\infty$ one gets 
\begin{gather}
I^{\mathrm{stop}}+\cc=H(M-K)L(x,t)+O(t^{-1}),
\label{c-trapped}
\\
L(x,t)=
\frac{\sqrt{1-\Omega_0^2}\,
\mathrm{e}^{-\sqrt{1-\Omega_0^2}|x|}}{\Omega_0(M\sqrt{1-\Omega_0^2}+1)}\,
\sin\Omega_0 t,
\label{S-MS-sol}
\end{gather}
see \cite{Gavrilov2022jsv,Gavrilov2019asm} and \ref{App-B}.
Here $L(x,t)$ is the localized non-vanishing oscillation,
$H$ is the Heaviside function.

In this paper we are mostly interested in the evaluation of the integral
$I^{\mathrm{pass}}$, which describes the propagating part of the wave-field.
Following to \cite{Gavrilov2022ijhmt,arxiv2206.08079},
we estimate it on a moving at an arbitrary sub-critical speed $w$ point of
observation. This approach is known for us due to \cite{Slepyan1972}. Taking
into account that solution 
\eqref{u-exact} clearly is an even function of $x$, put
\begin{gather}
|\xi|=w\tau,
\label{xi-w}
\\
0<w<1
\label{w-ineq}
\end{gather}
in the expression for $I^{\mathrm{pass}}$. The obtained integral can be
estimated for $t\to\infty$ 
using the method of stationary phase. This yields (see~\ref{App-A} for details)
\begin{gather}
I^\mathrm{pass}+\cc=-\frac{{A}(w)}{\sqrt t}\cos  \Big(\sqrt{1-w^2}\,\tau+\frac{\pi}{4}+\psi \Big)
+
O(\tau^{-3/2}),
\label{sum-I_1+I_2-single-wave}
\\
{A}(w)=\frac{\sqrt2 
w(1-w^2)^{1/4}H(1-w)}{\sqrt{\pi}\sqrt{4w^2(1-w^2)+(M-K+K w^2)^2}},
\label{sum-I_1+I_2-single-wave-amp}
\\
\psi =\arctan\frac{2w(1-w^2)^{1/2}}{M-K+Kw^2}.
\label{psi}
\end{gather}
Note that for $w>1$
there is no stationary point and the corresponding term of order $t^{-1/2}$ is
zero, this is taken into account by introducing the multiplier $H(1-w)$ in
the numerator of the right-hand side of 
Eq.~\eqref{sum-I_1+I_2-single-wave-amp}. 
{One can see that formulae 
\eqref{sum-I_1+I_2-single-wave}--\eqref{psi} describe the wave-field quite
well in the case $K>M$, when the trapped mode does not exist, see
Fig.~\ref{aloc-space.pdf}.} 
In the case $K<M$, when the trapped mode exists, we additionally should take
into account contribution 
from the trapped mode frequency $\Omega_0\in\mathbb S$, 
see Eqs.~\eqref{c-trapped}, \eqref{S-MS-sol} and
Fig.~\ref{loc-space.pdf}.
The numerical solution of the problem for Eq.~\eqref{OSC-maineq-SPRING}
presented in Figs.~\ref{aloc-space.pdf}, \ref{loc-space.pdf}
is obtained using an approach 
\cite{Gavrilov2019nody,Gavrilov2019asm}
based on finite difference schemes.
\begin{figure}[p]
\centering\includegraphics[width=0.85\textwidth]{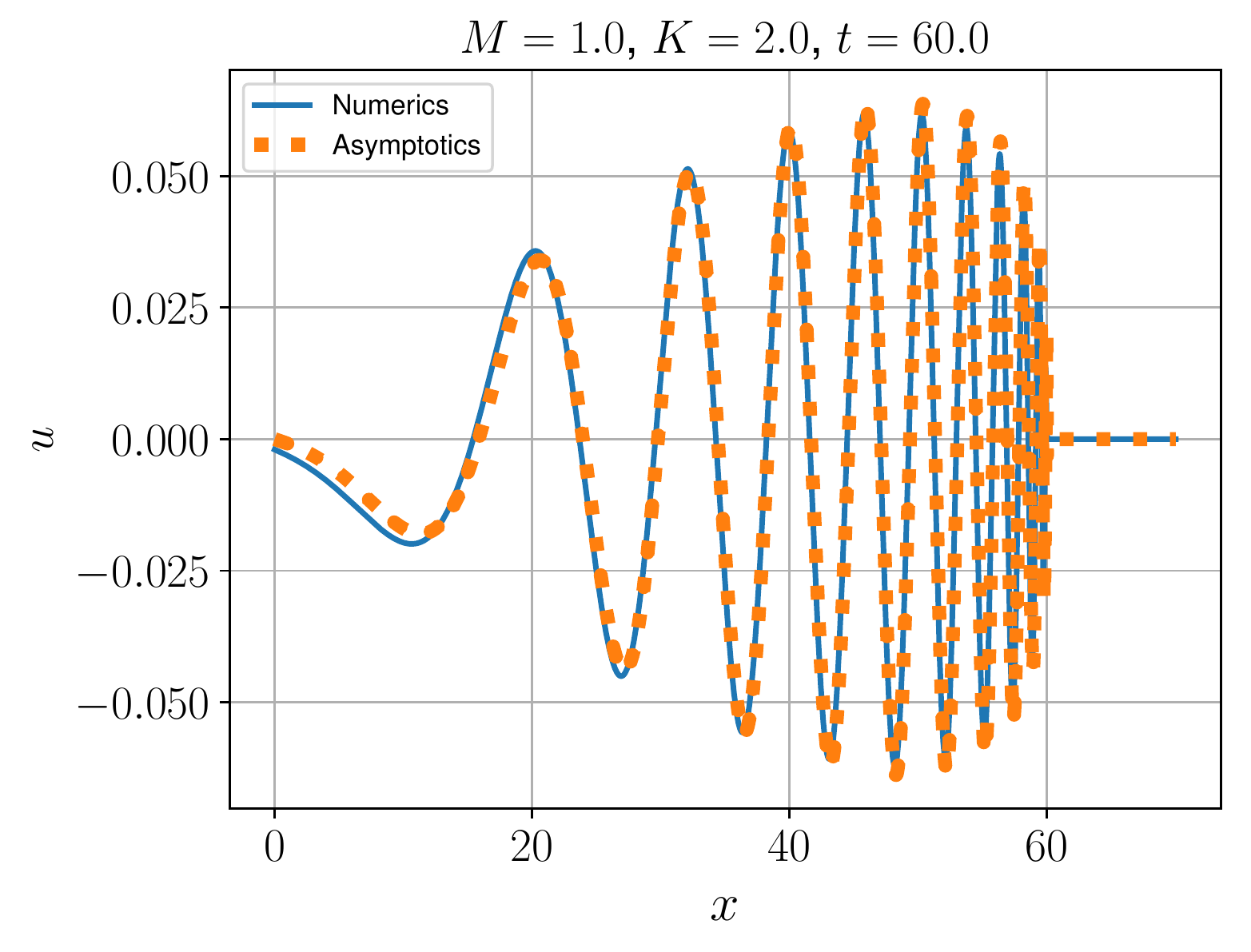}
\caption{Comparing of the asymptotic solution for $u$ 
given by Eqs.~\eqref{sum-I_1+I_2-single-wave}--\eqref{psi}, wherein $w=|x|/t$,
and the corresponding numerical solution of Eq.~\eqref{OSC-maineq-SPRING} in
the case $K>M$. One can observe the anti-localization near $x=0$.}
\label{aloc-space.pdf}
\end{figure}
\begin{figure}[p]
\centering\includegraphics[width=0.85\textwidth]{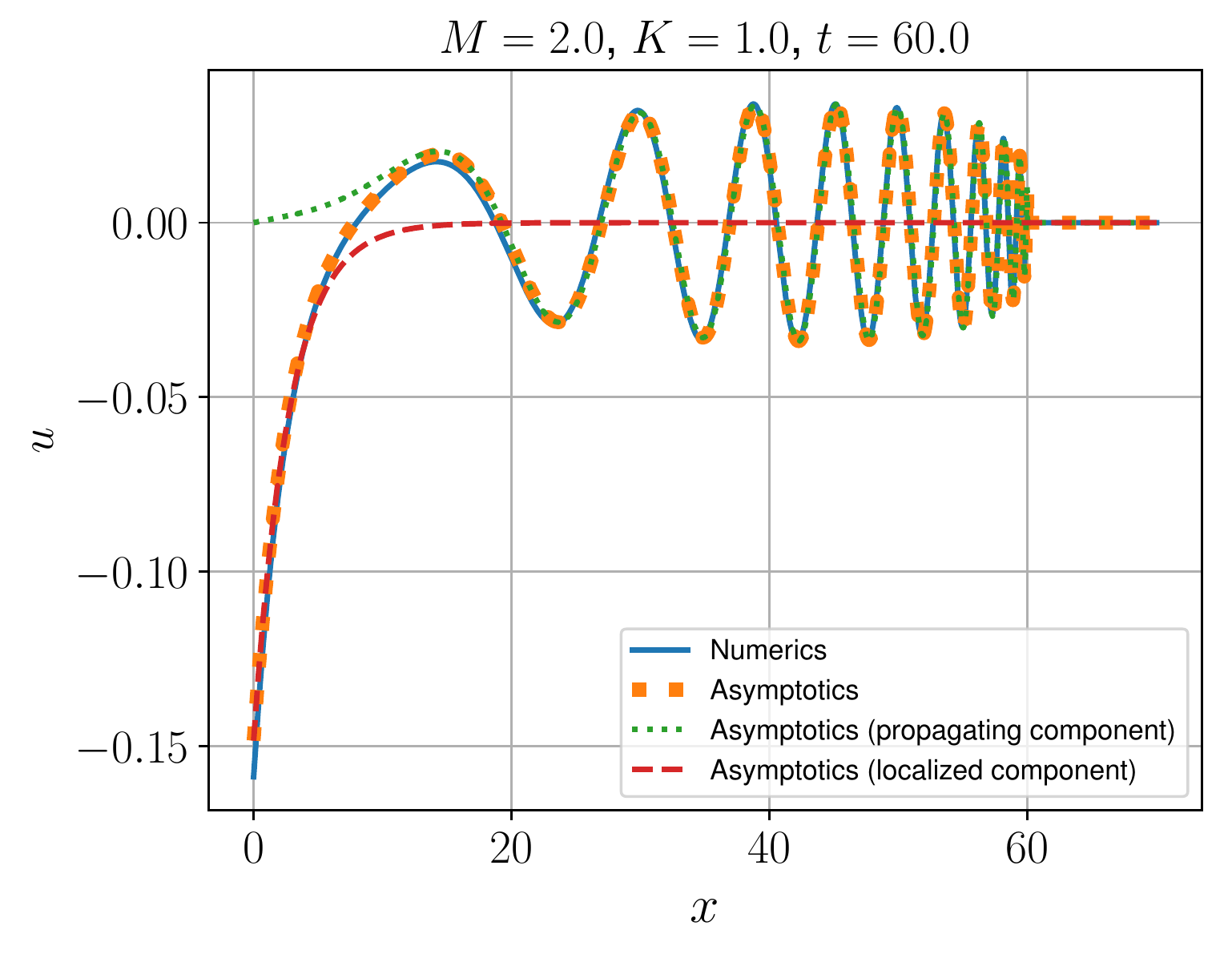}
\caption{Comparing of the asymptotic solution for $u$ 
given by Eqs.~\eqref{OSC-fr-root}--\eqref{psi}, wherein $w=|x|/t$,
and the corresponding numerical solution of Eq.~\eqref{OSC-maineq-SPRING} in
the case $K<M$. The anti-localization near $x=0$ co-exists with the
localization, see plots for the propagating component 
\eqref{sum-I_1+I_2-single-wave}
and the localized one
\eqref{S-MS-sol}.}
\label{loc-space.pdf}
\end{figure}


%


For $K\neq M$, i.e., everywhere in the parameter space excepting the boundary of
the localization domain \eqref{loc-domain}, the 
term of order $t^{ - 1 / 2}$ in expansion of $I^{\mathrm{pass}}$ is zero
at $w = 0$ (or $x=0$):
\begin{equation}
{A}(w)=\frac{\sqrt 2 w}{\sqrt{\pi}|K-M|}+O(w^3),\quad w\to0;\qquad A(0)=0;
\label{A-as}
\end{equation}
and, thus, the amplitude $A(w)$ 
of the propagating part $I^{\mathrm{pass}}+\cc$
for the string
displacements is small in a certain
neighbourhood of zero. From the physical point of view, this means that 
the amplitude of the  propagating component for the string
displacement (as well as the particle velocity or the strain) is
small in a certain expanding (since $w=|x|/t$)
neighbourhood of the inclusion.
{\it We call this phenomenon the anti-localization of non-stationary linear
waves.}
The greater the quantity
$|K-M|\neq0$, 
the wider the anti-localization zone for the propagating component
of the wave-field, and more energy concentrates closer to the leading wave-fronts
(see Fig.~\ref{slow.pdf}).
\afterpage{\clearpage}
\begin{figure}[htbp]
\centering\includegraphics[width=0.85\textwidth]{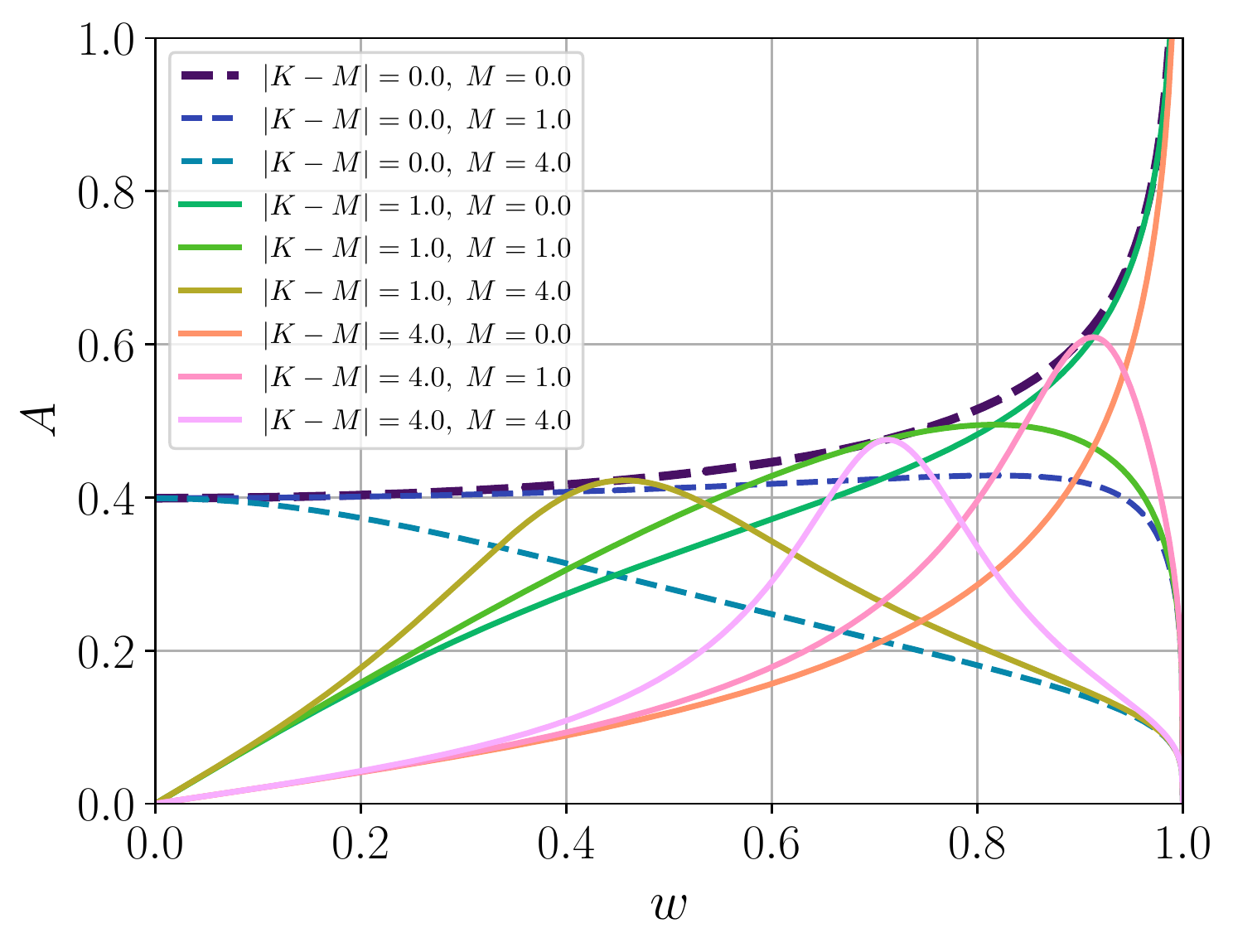}
\caption{The amplitude $A(w)$ defined by 
Eq.~\eqref{sum-I_1+I_2-single-wave-amp} for various system parameters (note
that $A(1)\to\infty$ if $M=0$ and $A(1)=0$ otherwise.)
In the case $K\neq M$ (see the solid lines) one can observe the anti-localization near $x=0$.}
\label{slow.pdf}
\end{figure}
The asymptotic expansion for the right-hand side of 
Eq.~\eqref{u-exact} {at $x=0$ (just at the
inclusion)} has the following form
\begin{equation}
u(0,t)=H(M-K)L(0,t)+\frac{2\sqrt{2}\,\cos{(t + \frac{\pi}{4})}}{\sqrt{\pi}\,(K-M)^2 \,t^{3/2}}
+ o(t^{-3/2}).
\label{S-MS-int2-fin}
\end{equation}
The first term  in the right-hand side of 
Eq.~\eqref{S-MS-int2-fin} is the contribution from the poles
$\pm\Omega_0$ and describes localized oscillation, which exists if and only
if $K<M$. 
The second term is the anti-localized part of the wave-field at the inclusion
expressed as 
the total contribution from the cut-off frequency
$\Omega_\ast$ for both integrals $I^{\mathrm{stop}}$ and $I^{\mathrm{pass}}$
(see \ref{App-C}).
{In Fig.~\ref{aloc-time.pdf} we compare the asymptotic solution given by 
Eq.~\eqref{S-MS-int2-fin} and corresponding numerical solution  of Eq.~\eqref{OSC-maineq-SPRING} in the case
$K>M$ and demonstrate a good agreement.}
\begin{figure}[htpb]
\centering\includegraphics[width=0.85\textwidth]{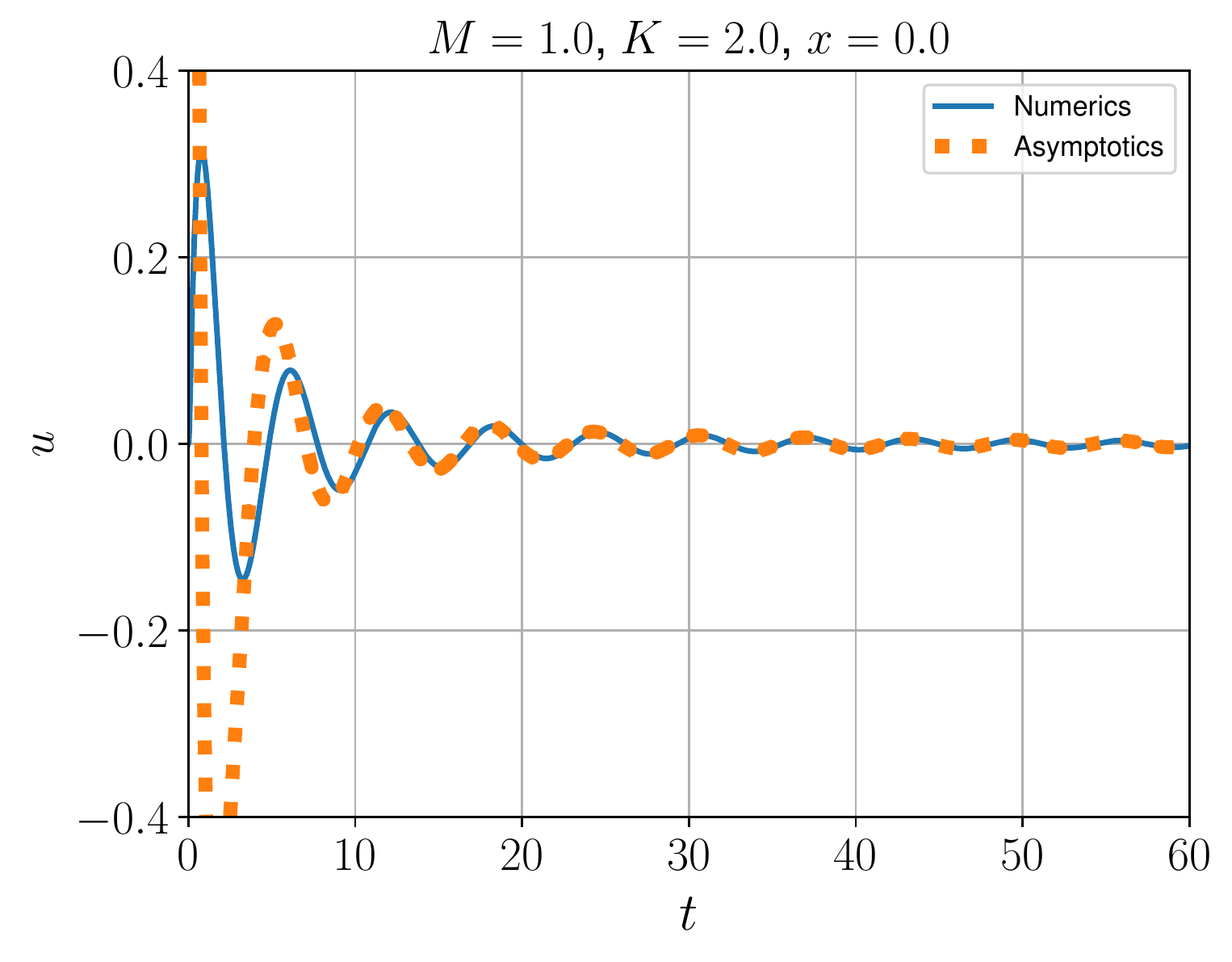}
\caption{Comparing the asymptotic solution for $u$ just 
at the inclusion in the form of the
right-hand side of Eq.~\eqref{S-MS-int2-fin}, and the corresponding numerical
solution of Eq.~\eqref{OSC-maineq-SPRING} in the case $K>M$.}
\label{aloc-time.pdf}
\end{figure}

\begin{remark}  
It is generally expectable that in the case of an impulse loading the vanishing oscillation at the source point 
(as well as at the inclusion position) is described by the contribution from a
cut-off frequency, since zero group velocity generally corresponds to such a
frequency \cite{hemmer1959dynamic,slepyan1987energy}. 
Thus, the disturbances, to which the cut-off
frequency corresponds, are accumulated near a source, whereas all other
disturbances are run away.
In this sense the presence of the cut-off frequency is very important feature
of the system under consideration. We expect that the behaviour of the system
without a cut-off frequency, where there is no accumulation of the
disturbances near a source, is quite different from one considered in our
paper.
\label{remark-slepyan}
\end{remark}

At the boundary of the localization domain $K=M$,  {one has} instead of 
Eq.~\eqref{A-as}:
\begin{equation}
{A}(w)=\frac{1}{\sqrt{2\pi}}+O(w^{2}),\quad w\to0.
\label{A0}
\end{equation}
This is the only case when the propagating wave-field defined by the integral
$I^{\mathrm{pass}}$ is not anti-localized.
\begin{figure}[htbp]
\centering\includegraphics[width=0.85\textwidth]{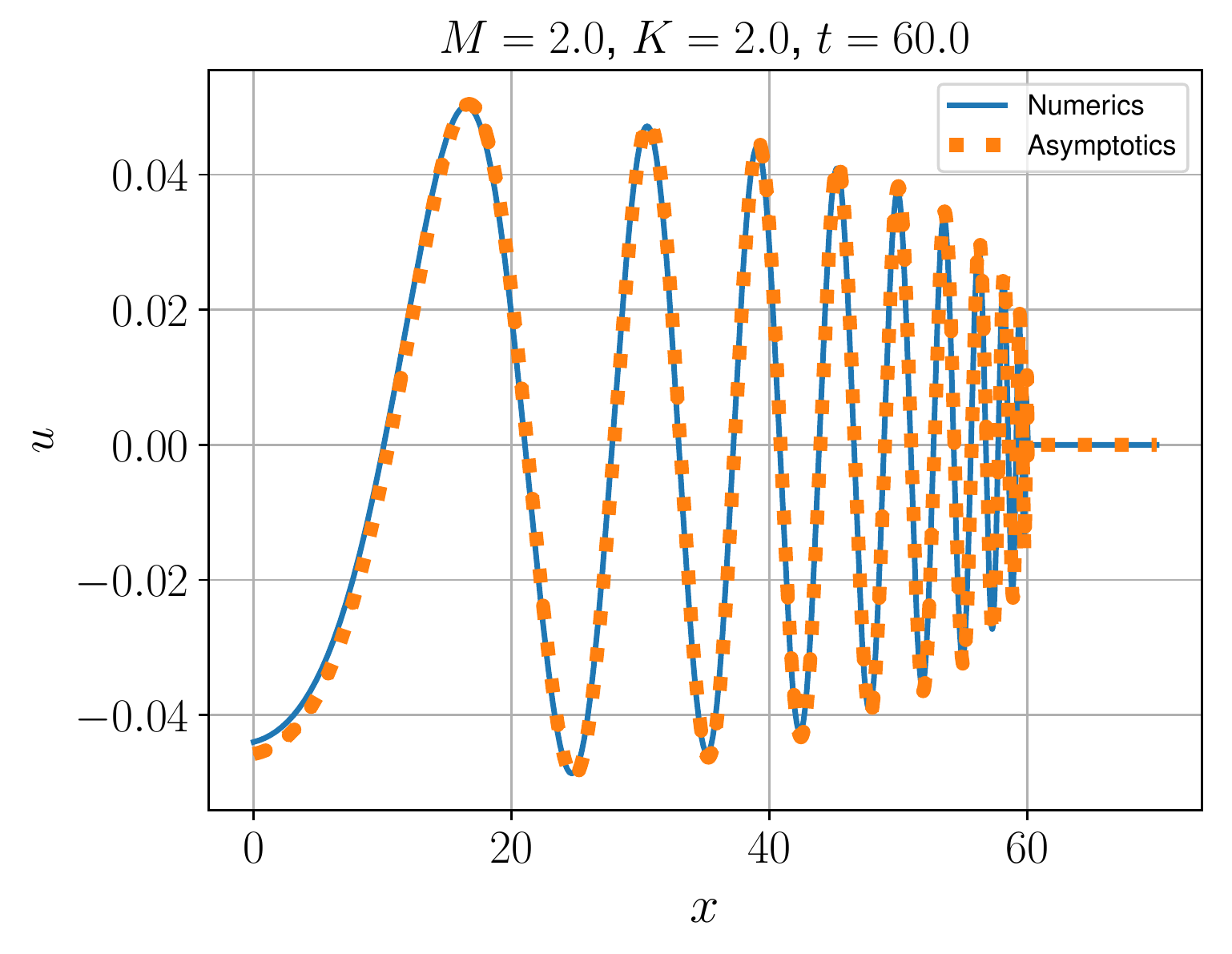}
\caption{Comparing of the asymptotic solution for $u$ 
given by  Eqs.~\eqref{sum-I_1+I_2-single-wave}--\eqref{psi}, wherein $w=|x|/t$,
and the corresponding numerical solution of Eq.~\eqref{OSC-maineq-SPRING} in
the case $K=M\neq0$. There is no anti-localization near $x=0$.}
\label{equal-space.pdf}
\end{figure}
The wave pattern, which corresponds to the case $K=M\neq0$, is shown in
Fig.~\ref{equal-space.pdf}.




\afterpage{\clearpage}
\section{Discussion}
\label{sec-discussion}
In the framework of the simple illustrative problem considered in the paper, we have
demonstrated the existence of the wave phenomenon, which we call the
anti-localization of non-stationary linear waves. 
The physical meaning of this phenomenon is
zeroing, or, strictly speaking, asymptotic weakening of the non-stationary 
wave-field in a neighbourhood 
of a discrete inclusion. The anti-localization is caused by a destructive
interference of the propagating
harmonics involved into the representation of the
solution in the form of Fourier integral $I^\mathrm{pass}$, see
\eqref{u-exact}.
Namely, the amplitude of the propagating component for the
string displacement $I^{\mathrm{pass}}+\cc$ is small in a certain expanding
with time
neighbourhood of the inclusion located at $x=0$, see Eq.~\eqref{A-as}  and the subsequent explanations.
Also, we have 
shown that the
observability of the anti-localization is deeply related with the phenomenon
of the localization, since the anti-localization exists for all cases excepting the
boundary $(K=M)$ of the localization domain \eqref{loc-domain}.
For a system, which
parameters lie in the parameter space outside the localization domain ($K>M$), 
the anti-localization can be easily discovered in plots characterizing spatial
distribution of a non-stationary wave-field
(compare Fig.~\ref{aloc-space.pdf} with Fig.~\ref{equal-space.pdf}).
{For a system, which parameters lie inside a
localization domain ($K<M$) the anti-localization co-exists with the
localization, 
making the anti-localization difficult to recognize in plots.
In the latter case, one can see in Fig.~\ref{loc-space.pdf} that the propagating
component is anti-localized, whereas the total solution demonstrates a
localized behaviour.
{We have demonstrated how the anti-localization influences on the propagating
part of the wave-field at an arbitrary spatial position, see 
Eqs.~\eqref{xi-w}--\eqref{psi} and Fig.~\ref{slow.pdf}.

For the best of our knowledge, the anti-localization of non-stationary waves was
never treated before as a general wave phenomenon, but our results are in
agreement with particular observations in studies, where the
non-stationary oscillation caused by an impulse source at an inclusion in some
infinite (discrete
\cite{hemmer1959dynamic,Mueller1962,Kashiwamura1962,Rubin_1963,Mueller2012} or continuum \cite{kaplunov1986torsional}) systems 
have been considered.  In all of these systems under certain conditions
trapped modes can be observed.
Namely, 
in \cite{hemmer1959dynamic,Mueller1962,Kashiwamura1962,Rubin_1963,Mueller2012},
as well as in our study \cite{arxiv2206.08079}, a 1D harmonic chain with
an isotopic defect is under consideration. In \cite{kaplunov1986torsional} the problem
mathematically equivalent to our illustrative one with additional constraint $K=0$ is
investigated. In all studies
\cite{kaplunov1986torsional,hemmer1959dynamic,Mueller1962,Kashiwamura1962,Rubin_1963,Mueller2012}
pure inertial inclusions were considered, and
the contributions from the cut-off frequency, describing vanishing oscillation 
just at an inclusion,  were obtained.
These results are analogues to our
Eq.~\eqref{S-MS-int2-fin} with the same order of vanishing $t^{-3/2}$ of the
solutions. In 
\cite{Kashiwamura1962,Rubin_1963} it is indicated that without a defect the
order of vanishing is $t^{-1/2}$, see our Eqs.~\eqref{sum-I_1+I_2-single-wave}, 
\eqref{A0} wherein $w=0$.
At the same time, the alteration of the propagating component of the wave-field
at an arbitrary spatial position due to the presence of an inclusion
in \cite{kaplunov1986torsional,hemmer1959dynamic,Mueller1962,Kashiwamura1962,Rubin_1963,Mueller2012}
remains to be unstudied.
For our problem it is described by
Eqs.~\eqref{xi-w}--\eqref{psi}.

{Note that commonly while investigating the systems, where the
localization is possible, the authors estimate analytically only the
contribution from the stop-band related with the localization. The
corresponding  vanishing contribution from the pass-band often remains in the
integral form (see, for example, \cite{Mishuris2020,Yu2019}) and only a few
attempts
\cite{kaplunov1986torsional,hemmer1959dynamic,Mueller1962,Kashiwamura1962,Rubin_1963,Mueller2012}
to estimate it analytically have been done. Apparently, the vanishing waves
from the pass-band are considered as the less interesting and important part
of the solution in the presence of the localized modes, and, therefore, it is
easier to calculate it numerically.}


The results of 
\cite{kaplunov1986torsional,hemmer1959dynamic,Mueller1962,Kashiwamura1962,Rubin_1963,Mueller2012} 
were obtained in the case,
where the boundary of the localization domain corresponds to a homogeneous
system without any inclusion ($M=0$, $K=0$ for our illustrative problem). 
{Thus, in this paper we first time have demonstrated that the emergence and the intensity
of the anti-localization in a system are related not with its non-uniformity
itself, but
with the position in the parameter space outside the boundary 
$K=M$ of the localization domain}. 
Indeed, the analysis of Eqs.~\eqref{sum-I_1+I_2-single-wave-amp}, \eqref{A-as}
shows that the greater the quantity
$|K-M|\neq0$, the wider the anti-localization zone for the
propagating component of the wave-field, and more wave energy concentrates closer to
the leading wave-fronts
(see the solid lines in Fig.~\ref{slow.pdf}).

Note that in
\cite{hemmer1959dynamic,Mueller1962,Kashiwamura1962,Rubin_1963,Mueller2012}
stochastic processes of heat transfer are
considered, thus, for a reader, who is not familiar with such kind of problems, it is
not so easy to understand that the discussed effects have a deterministic
nature. On the other hand,  in \cite{kaplunov1986torsional} the 
anti-localization always co-exists with the localization, and therefore it is not
considered as an important effect. 



{The observability of the anti-localization and its relation to the localization in 
systems with more than one stop-bands and cut-off frequencies,\footnote{Recall that the behaviour of the
anti-localized part of the wave-field just at the
inclusion, i.e., of the second term in the right-hand side of
Eq.~\eqref{S-MS-int2-fin},
is determined by the total contribution from the cut-off frequency
for both integrals $I^{\mathrm{stop}}$ and $I^\mathrm{pass}$, see also
Remark~\ref{remark-slepyan}.}
as well as in systems characterized by more than one dispersion curves,
2D--3D systems, systems with distributed inclusions
remain to be the open questions.} 
The case when a source and an inclusion
are located
at the different positions also requires an additional investigation. The
possible practical application of the phenomenon is an acoustical isolation
from waves caused by 
impulse and stochastic loadings and, in particular, seismic protection.
{Apparently, the anti-localization that we discuss also can be observed 
in stationary ordered stochastic systems since a loading in the form of a
white noise contains harmonics with all possible frequencies.
This can probably be an explanation for the phenomenon of ``a
cold point'' discussed in \cite{Paul2020,Gendelman2021}.}

\section*{Acknowledgements}
The authors are grateful to A.P.~Kiselev, S.A.~Kukushkin,
\framebox{D.A.~Indeisev}, O.V.~Gendelman, O.V.~Motygin,
A.M.~Kriv\-tsov, V.A.~Kuzkin, {A.A.~Sokolov}
for useful and stimulating discussions.
 
\section*{Declaration of competing interest}
None to declare.

\section*{Funding}
This work is supported by Russian Science Support Foundation (project 22-11-00338).

\appendix

\section{Calculation of the asymptotics for the integral $I^{\mathrm{pass}}$}
\label{App-A}
The technique of asymptotic evaluation of the integral $I^\mathrm{pass}$ in the
case of a string on elastic foundation is quite similar to one used in
\cite{arxiv2206.08079} in the case of a discrete chain with an isotopic defect. 
For the reader's convenience we provide here the corresponding calculations.

According to the method of stationary phase \cite{Fedoryuk1977,temme2014}
the large time asymptotics of a Fourier integral
\begin{equation}
I=\int_a^b
\mathcal A
(\Omega)\,{\EXP{\I\phi(\Omega)t}\,\d\Omega}
\end{equation}
is the sum of 
contributions $I(\Omega_i)$
\begin{gather}
I(\Omega_i)\=
\int_a^b \mathscr A(\Omega)\,\chi_{\Omega_i}(\Omega)\,\EXP{\I\phi(\Omega)t} \,\d\Omega
\label{contrib}
\end{gather}
from all the critical points $\Omega_i$ inside the integration interval $[a,b]$:
\begin{equation}
I=\sum_i I(\Omega_i)+O(t^{-\infty}).
\end{equation}
The critical points are stationary points for the phase $\phi(\Omega)$, 
finite end-points of
the integration intervals and points of non-smoothness for the phase
$\phi(\Omega)$, or the 
amplitude $\mathscr A(\Omega)$, or their derivatives. Functions $\chi_{\Omega_i}(\Omega)$ defined for
every critical point $\Omega_i$ are
neutralisers\footnote{Neutraliser \cite{temme2014,van1948method} $\chi_{\Omega_i}(\Omega)$ at a
critical point $\Omega_i$ is an infinitely differentiable function such that
$\chi_{\Omega_i}(\Omega_i)=1$, $\chi^{(n)}_{\Omega_i}(\Omega_i)=0$ for $n>1$,
and $\chi_{\Omega_i}(\Omega)\equiv0$ outside of a certain neighbourhood of
$\Omega_i$.}
at $\Omega=\Omega_i$ such that
$\chi_{\Omega_i}(\Omega)\equiv0$ in a neighbourhood of any $\Omega_j$
for $j\neq i$. Introducing of neutralisers allows one to calculate the
contribution from a critical point separately.

The integral $I^\mathrm{pass}$
at the moving point
of observation \eqref{xi-w} can be rewritten as follows:
\begin{gather}
I^\mathrm{pass}=\frac1{2\pi}\int_1^\infty
\mathcal A
^\mathrm{pass}
(\Omega)\,{\EXP{\I\phi(\Omega)t}\,\d\Omega},
\\
\mathcal A^\mathrm{pass}(\Omega)=\frac{-1}{2\I\sqrt{\Omega^2-1}-K+M\Omega^2},
\\
\phi(\Omega)=\sqrt{\Omega^2-1}w-\Omega.
\label{fi-1}
\end{gather}
The critical points for $I^\mathrm{pass}$ are the stationary
point for $\phi$ where 
\begin{equation}
\phi'_\Omega=0, 
\label{phases-prime}
\end{equation}
and the singular end-point $\Omega=\Omega_\ast=1$. 
One has
\begin{gather}
\begin{multlined}       
\mathscr A^\mathrm{pass}(\Omega)
=
\mathscr A^\mathrm{pass}_0
+
\mathscr A^\mathrm{pass}_{1/2}\sqrt{\Omega-1}+O(\Omega-1)
\\\qquad=\frac1{K-M}+\frac{2\I\sqrt2}{(K-M)^2}\sqrt{\Omega-1}+O(\Omega-1),
\end{multlined}
\label{A-pass-expa}
\\
\phi(\Omega)=-1+\sqrt2w\sqrt{\Omega-1}-(\Omega-1)+o(\Omega-1).
\label{phi-pass-expa}
\end{gather}
Using the last formulae and applying the Erd\'elyi lemma (see \ref{sec-erdelyi})
wherein $\alpha=1/2,\ \beta=1$,
one can show that the contribution from the end-point
$\Omega=\Omega_\ast\equiv1$ is $O(t^{-2})$ if 
\eqref{w-ineq} is fulfilled.

Resolving Eq.~\eqref{phases-prime} shows that if \eqref{w-ineq} is
true, then there exists a unique non-degenerate stationary point:
\begin{gather}
\Omega_{\mathrm{s}}=\frac1{\sqrt{1-w^2}},
\\
\phi(\Omega_\mathrm{s})=-\sqrt{1-w^2},
\\
\phi''(\Omega_\mathrm{s})=-\frac{(1-w^2)^{3/2}}{w^2}<0.
\label{phase1-second-deriv}
\end{gather}

Calculating the contribution from a stationary point 
\cite{erdelyi1956asymptotic,Fedoryuk1977}
yields the following asymptotics for integrals $I^\mathrm{pass}$:
\begin{multline}
I^\mathrm{pass}(\Omega_\mathrm{s})=
\frac{\mathscr A^\mathrm{pass}(\Omega_\mathrm{s})}{2\pi}\sqrt{\frac{2\pi}{|\phi''(\Omega_\mathrm{s})|t}}
\EXP{\I\phi(\Omega_\mathrm{s})t+\frac {\I\pi}4\sign \phi''(\Omega_\mathrm{s})}
+O(t^{-3/2})
\\=
-\frac1{\sqrt{2\pi t}}\frac{w(1-w^2)^{1/4}
\,
\mathrm e^{-\I\sqrt{1-w^2}\tau-\frac{\I\pi}{4}}
}
{2\I w\sqrt{1-w^2}+M-K(1-w^2)}+O(\tau^{-3/2}).
\label{I_1-st-phase-K=const}
\end{multline}
Thus,
\begin{multline}
I^\mathrm{pass}+\cc
=\frac{
w(1-w^2)^{1/4}
}
{\sqrt{2\pi t\,}\big(4w^2(1-w^2)+(M-K(1-w^2))^2\big)}
\\
\times\bigg(   
-2(M
-K(1-w^2))\cos \Big (\sqrt{1-w^2}\tau+\frac{\pi}{4}\Big )
+
\\
4w\sqrt{1-w^2}\sin \Big(\sqrt{1-w^2}\tau+\frac{\pi}{4}\Big)
\bigg)
+
O(\tau^{-3/2}).
\label{sum-I_1+I_2-Re-K=const}
\end{multline}
The last formula can be transformed to the form of
Eqs.~\eqref{sum-I_1+I_2-single-wave}--\eqref{psi}.

\section{Calculation of the asymptotics for the integral $I^{\mathrm{stop}}$}
\label{App-B}
One has
\begin{gather}
I^\mathrm{stop}=
\frac1{2\pi}\int_0^1 
\frac{\EXP{-\sqrt{1-\Omega^2}|\xi|-\I\Omega t}}{2\sqrt{1-\Omega^2}+K-M\Omega^2}
\, \d\Omega
=
\frac1{2\pi}\int_0^1 
\mathscr A^\mathrm{stop}(\Omega)\,
\EXP{-\I\Omega t}
\, \d\Omega
,
\\
\mathscr A^\mathrm{stop}(\Omega,x)\=
\frac{\EXP{-\sqrt{1-\Omega^2}|\xi|}}{2\sqrt{1-\Omega^2}+K-M\Omega^2}.
\end{gather}
The denominator of $\mathscr A^\mathrm{stop}(\Omega,x)$ is the frequency
equation \cite{Gavrilov2022jsv,Gavrilov2019asm} for a trapped mode, which exists in the system if and only if $K<M$:
\begin{equation}
{2\sqrt{1-\Omega^2}+K-M\Omega^2}=0.	
\label{freq-eq}
\end{equation}

In the case $K>M$ there are two critical points for integral $I^\mathrm{stop}$,
namely, the finite end-points $\Omega=0$ and $\Omega=\Omega_\ast=1$.
The contribution $I^\mathrm{stop}(0)$ from the end-point $\Omega = 0$
totally compensates by the complexly conjugated integral over $( - 1 , 0)$,
see term $\cc$ in Eq.~\eqref{u-exact}.
For $\Omega\to1-0$ one obtains:  
\begin{multline}
\mathscr A^\mathrm{stop}(\Omega,x)=
\mathscr A^\mathrm{stop}_0
+
\mathscr A^\mathrm{stop}_{1/2}(x)\sqrt{1-\Omega}+O(1-\Omega)
\\=
\frac{1}{K-M}-\frac{\sqrt{2} \big((K -M)
|x|+2\big)}{(K-M)^2}\sqrt{1-\Omega}+O(1-\Omega).
\label{A-stop-expa}
\end{multline}
Now, applying 
the Erd\'elyi lemma (see \ref{sec-erdelyi}) wherein $\alpha=1,\ \beta=1$,
we can estimate the contribution from the end-point $\Omega=1$ 
as $O(t^{-1})$. 

In the case $K<M$
there exists a
{simple root~$\Omega_0$} of 
\eqref{freq-eq}
and, therefore,
integral $I^{\mathrm{stop}}$ does not exist in the classical
sense.
In the
latter case,
integral $I^{\mathrm{stop}}$ 
should be considered as the Fourier transform for
a generalized function. To regularize this {we can apply} 
\cite{Gavrilov1999jsv,arxiv2206.08079}
the limit absorption
principle and add the term $-2\gamma \dot u,\ \gamma>0$ describing the viscous
friction into the left-hand side of governing equation 
\eqref{OSC-maineq-SPRING}.
The frequency equation \eqref{freq-eq} transforms to 
\begin{equation}
{2\sqrt{1-\Omega^2-2\gamma\I\Omega}+K-M\Omega^2}=0.	
\label{freq-eq-mod}
\end{equation}
Accordingly, one can check that the root satisfying
Eq.~\eqref{OSC-fr-root},
transforms into $\Omega_0-\I0$ as $\gamma\to+0$. Thus, calculating the
contribution $I^\mathrm{stop}(\Omega_0)$ one gets 
\cite{Fedoryuk1977,arxiv2206.08079,Gavrilov2019asm}:
\begin{multline}
I^\mathrm{stop}(\Omega_0)=\frac{H(M-K)}{2\pi}\int_0^1
\chi_{\Omega_0}(\Omega)
\left(
\frac{\Res\big(\mathscr A^\mathrm{stop},\Omega_0\big)}
{\Omega-\Omega_0+\I0}
+O(1)
\right)
\EXP{-\I\Omega t}
\,\d\Omega
\\+O(t^{-\infty})
=
-\I H(M-K)\Res\big(\mathscr A^\mathrm{stop},\Omega_0\big)\,\EXP{-\I\Omega_0 t}
+O(t^{-\infty})
\\=-\frac{H(M-K)\sqrt{1-\Omega_0^2}\,
\mathrm{e}^{-\sqrt{1-\Omega_0^2}|x|}}{2\I\Omega_0(M\sqrt{1-\Omega_0^2}+1)}\,
\EXP{-\I\Omega_0 t}
+O(t^{-\infty})
.
\label{Istop-W0}
\end{multline}
Here symbol $\Res (f, \Omega_0)$ means the residue of a function $f
(\Omega)$ at a pole $\Omega =\Omega_0$, $H$ is the Heaviside function.
Finally, calculating $I^\mathrm{stop}+\cc$ leads to formulae~
\eqref{c-trapped}, \eqref{S-MS-sol}.

An alternative approach 
\cite{Mishuris2020,kaplunov1986torsional,Rubin_1963} to calculate the
contribution from the stop-band,
which is more widespread in the literature, is to use the Laplace transform
instead of the Fourier transform for generalized functions, and 
to modify integration path to a closed one
according to the Jordan lemma
using
{branch cuts}. In the framework of the latter approach a
trapped mode (if exists) can be taken into account by the residue theorem.
The integral over the branch cuts should be estimated by the
{Erd\'elyi} lemma. 
To use the alternative approach one needs to introduce the corresponding
alterations into the calculations in \ref{App-C}.
Both approaches lead to the identical results.

\section{Calculation of the asymptotics for the inclusion displacements in the
case $K\neq M$}
\label{App-C}

Here we derive asymptotic formula
\eqref{S-MS-int2-fin}.
One has:
\begin{equation}
u(0,t)=I^\mathrm{stop}(\Omega_0)\big|_{x=0}
+I^\mathrm{stop}(\Omega_\ast)\big|_{x=0}
+I^\mathrm{pass}(\Omega_\ast)\big|_{w=0}.
\end{equation}
The first term in the right-hand side of the last formula is given by 
Eqs.~\eqref{c-trapped},
\eqref{S-MS-sol}.
Now we should calculate the second and the third terms.

Since, according to Eqs.~\eqref{A-pass-expa}, \eqref{A-stop-expa},
$\mathscr A_0^\mathrm{pass}=\mathscr A_0^\mathrm{stop}$ 
\begin{equation}
\underbrace{\int_{1}^{\infty}
\chi_1(\Omega)\mathscr A_0^\mathrm{pass}\,\EXP{-\I\Omega t}
\,\d \Omega}_{J}
+
\int_{0}^{1}
\chi_1(\Omega)
\mathscr A_0^\mathrm{stop}\,\EXP{-\I\Omega t}
\,\d \Omega=O(t^{-\infty}).
\end{equation}
Applying the 
{Erd\'elyi} lemma wherein $\alpha=1$, $\beta=3/2$, one gets 
\begin{multline}
I^\mathrm{pass}(\Omega_\ast)\big|_{w=0}+\cc
\\=
J+\frac\I{2\pi}{\int_{1}^{\infty}
\chi_1(\Omega)
\big(\big|\mathscr A_{1/2}^\mathrm{pass}\big|\sqrt{\Omega-1}+o(\sqrt{\Omega-1})\big)
\,\EXP{-\I\Omega t}
\,\d \Omega}+\cc
\\=
J+
\frac\I{2\pi}\int_0^{\infty} 
\chi_1(\mu+1)
\big(\big|\mathscr A^\mathrm{pass}_{1/2}\big|\sqrt\mu
+o(\sqrt\mu)
\big)
\EXP{-\I(\mu+1)t}\,\d\mu+\cc+O(t^{-\infty})
\\=
2\Re J+2\Re
\frac{\big|\mathscr A^\mathrm{pass}_{1/2}\big|
\Gamma\left(\frac32\right)\EXP{\I\left(\frac\pi2-\frac{3\pi}4-t\right)}}
{2\pi\, t^{3/2}}
+ o(t^{-3/2})
\\=
2\Re J+\frac{\sqrt2\cos\left(t+\frac{\pi}4\right)}{\sqrt\pi(K-M)^2\, t^{3/2}}
+ o(t^{-3/2}),
\label{Ipass1}
\end{multline}
\begin{multline}  
I^\mathrm{stop}(\Omega_\ast)\big|_{x=0}+\cc
\\=
-J-\frac1{2\pi}{\int_{0}^{1}
\chi_1(\Omega)
\big(\big|\mathscr A_{1/2}^\mathrm{stop}(0)\big|\sqrt{1-\Omega}+o(\sqrt{1-\Omega})\big)
\,\EXP{-\I\Omega t}
\,\d \Omega}+\cc
\\=
-J-\frac1{2\pi}{\int_{-1}^{0}
\chi_{1}(-\Omega)
\big(\big|\mathscr A_{1/2}^\mathrm{stop}(0)\big|\sqrt{\Omega+1}+o(\sqrt{\Omega+1})\big)
\,\EXP{\I\Omega t}
\,\d \Omega}+\cc
\\=
-J
-\frac1{2\pi}\int_0^1
\chi_1(1-\mu)
\big(\big|\mathscr A^\mathrm{stop}_{1/2}(0)\big|\sqrt\mu
+o(\sqrt\mu)\big)
\EXP{\I(\mu-1)t}\,\d\mu+\cc+O(t^{-\infty})
\\=
-2\Re J+
2\Re
\frac
{\big|\mathscr A^\mathrm{stop}_{1/2}(0)\big|
\Gamma\left(\frac32\right)\EXP{\I\left(\frac{3\pi}4-\pi-t\right)}}
{2\pi\, t^{3/2}}
+ o(t^{-3/2})
\\=
-2\Re J+
\frac{\sqrt2\cos\left(t+\frac{\pi}4\right)}{\sqrt\pi(K-M)^2\, t^{3/2}}
+ o(t^{-3/2})
.
\label{Istop1}
\end{multline}
Here $\Gamma$ is the Gamma function, $\Gamma(3/2)=\sqrt\pi/2$.
Now, formulae~\eqref{c-trapped},
\eqref{S-MS-sol},
\eqref{Ipass1},
\eqref{Istop1} result in Eq.~\eqref{S-MS-int2-fin}.

{Equivalently, the second term of Eq.~\eqref{S-MS-int2-fin}
can be obtained by applying
the Erd\'elyi lemma to calculate
the contribution from the branch cuts \cite{kaplunov1986torsional}.}
Both approaches lead to the identical results.

\section{The Erd\'elyi lemma}
\label{sec-erdelyi}
\begin{theorem} Let $a>0,\ \alpha\geq1,\ \beta>0$, $f(\Omega)\in C^\infty$,
$f^{(n)}(a)=0\ \forall n.$ 
Then
\begin{gather}
\int_0^a\Omega^{\beta-1}f(\Omega)\,\EXP{\I t\Omega^\alpha}\,\d\Omega\sim
\sum_{k=0}^\infty c_k t^{-\frac{k+\beta}\alpha},\quad t\to\infty;
\\
c_k=\frac{f^{(k)}(0)}{k!\alpha}\,
\Gamma\left(\frac{k+\beta}\alpha\right)\EXP{\frac{\I\pi(k+\beta)}{2\alpha}}.
\end{gather}
\end{theorem}
The proof can be found in \cite{erdelyi1956asymptotic,Fedoryuk1977}.

In
\ref{App-A}
we 
apply the Erd\'elyi lemma to an integral,
where $\beta=1$, $\alpha=1/2$. The corresponding asymptotics can be obtained
by taking $\Omega^\alpha$ as the new integration variable, and applying the
Erd\'elyi lemma to the obtained integral.

\providecommand*{\BibDash}{}
\providecommand*{\BibEmph}[1]{{\it #1}}
\bibliographystyle{elsarticle-num}
\biboptions{compress}
\bibliography{bib/serge-gost,bib/mode,bib/mode-trans,bib/math,bib/impurity,bib/discrete,bib/weak,bib/thermo,bib/embedded}

\end{document}